\documentclass[aps,prl,twocolumn,superscriptaddress,showpacs]{revtex4-1}

\usepackage{graphicx}

\begin{document}


\title{Direct Evidence of Octupole Deformation in Neutron-Rich $^{144}$Ba}


\author{B. Bucher}
\email[]{bucher3@llnl.gov}
\affiliation{Lawrence Livermore National Laboratory, Livermore,
California 94550, USA}
\author{S. Zhu}
\affiliation{Argonne National Laboratory, Argonne, Illinois 60439,
USA}
\author{C. Y. Wu}
\affiliation{Lawrence Livermore National Laboratory, Livermore,
California 94550, USA}
\author{R. V. F. Janssens}
\affiliation{Argonne National Laboratory, Argonne, Illinois 60439,
USA}
\author{D. Cline}
\author{A. B. Hayes}
\affiliation{University of Rochester, Rochester, New York 14627,
USA}
\author{M. Albers}
\author{A.~D.~Ayangeakaa}
\affiliation{Argonne National Laboratory, Argonne, Illinois 60439,
USA}
\author{P.~A.~Butler}
\affiliation{University of Liverpool, Liverpool L69 7ZE, United
Kingdom}
\author{C. M. Campbell}
\affiliation{Lawrence Berkeley National Laboratory, Berkeley,
California 94720, USA}
\author{M. P. Carpenter}
\affiliation{Argonne National Laboratory, Argonne, Illinois 60439,
USA}
\author{C. J. Chiara}
\altaffiliation[Present address: ]{U.S. Army Research Laboratory, 
Adelphi, Maryland 20783, USA}
\affiliation{Argonne National Laboratory, Argonne, Illinois 60439,
USA} \affiliation{University of Maryland, College Park, Maryland
 20742, USA}
\author{J.~A. Clark}
\affiliation{Argonne National Laboratory, Argonne,
Illinois 60439, USA}
\author{H.~L.~Crawford}
\altaffiliation[Present address: ]{Lawrence Berkeley National
Laboratory, Berkeley, California 94720, USA} \affiliation{Ohio
University, Athens, Ohio 45701, USA}
\author{M.~Cromaz}
\affiliation{Lawrence Berkeley National Laboratory, Berkeley,
California 94720, USA}
\author{H. M. David}
\altaffiliation[Present address: ]{GSI Helmholtzzentrum f\"{u}r
Schwerionenforschung, 64291 Darmstadt, Germany} \affiliation{Argonne 
National Laboratory, Argonne, Illinois 60439, USA}
\author{C. Dickerson}
\affiliation{Argonne National Laboratory, Argonne, Illinois 60439,
USA}
\author{E. T. Gregor}
\affiliation{University of the West of Scotland, Paisley PA1 2BE,
United Kingdom}
\affiliation{SUPA, Scottish Universities Physics Alliance, 
Glasgow G12 8QQ, United Kingdom}
\author{J. Harker}
\affiliation{Argonne National Laboratory, Argonne,
Illinois 60439, USA} \affiliation{University of Maryland, College
Park, Maryland 20742, USA}
\author{C.~R.~Hoffman}
\author{B. P.~Kay}
\author{F.~G.~Kondev}
\affiliation{Argonne National Laboratory, Argonne, Illinois 60439,
USA}
\author{A. Korichi} \affiliation{Argonne National Laboratory,
Argonne, Illinois 60439, USA} \affiliation{CSNSM, IN2P3-CNRS,
bat 104-108, F-91405 Orsay Campus, France}
\author{T. Lauritsen}
\affiliation{Argonne National Laboratory, Argonne, Illinois 60439,
USA}
\author{A. O. Macchiavelli}
\affiliation{Lawrence Berkeley National Laboratory, Berkeley,
California 94720, USA}
\author{R.~C. Pardo}
\affiliation{Argonne National Laboratory, Argonne, Illinois 60439,
USA}
\author{A.~Richard}
\affiliation{Ohio University, Athens, Ohio 45701, USA}
\author{M.~A.~Riley}
\affiliation{Florida State University, Tallahassee, Florida 32306,
USA}
\author{G. Savard}
\affiliation{Argonne National Laboratory, Argonne, Illinois 60439,
USA}
\author{M. Scheck}
\affiliation{University of the West of Scotland, Paisley PA1 2BE,
United Kingdom}
\affiliation{SUPA, Scottish Universities Physics Alliance, 
Glasgow G12 8QQ, United Kingdom}
\author{D. Seweryniak}
\affiliation{Argonne National Laboratory, Argonne, Illinois 60439,
USA}
\author{M. K. Smith}
\affiliation{University of Notre Dame, Notre Dame, Indiana 46556,
USA}
\author{R. Vondrasek}
\affiliation{Argonne National Laboratory, Argonne, Illinois 60439,
USA}
\author{A. Wiens}
\affiliation{Lawrence Berkeley National Laboratory, Berkeley,
California 94720, USA}


\date{\today}

\begin{abstract}
The neutron-rich nucleus $^{144}$Ba ($t_{1/2}$=11.5 s) is expected
to exhibit some of the strongest octupole correlations among nuclei with 
mass numbers $A$ less than 200.  
Until now, indirect evidence for such strong correlations has been
inferred from observations such as enhanced $E1$ transitions 
and interleaving positive- and negative-parity levels in the
ground-state band.  In this experiment, the octupole strength was
measured directly by sub-barrier, multi-step Coulomb excitation of a
post-accelerated 650-MeV $^{144}$Ba beam on a 1.0-mg/cm$^2$ $^{208}$Pb 
target. The measured value of the matrix element, 
$\langle 3_1^- \| \mathcal{M}(E3) \| 0_1^+ \rangle=0.65(^{+17}_{-23})$
$e$b$^{3/2}$, corresponds to a reduced $B(E3)$ transition probability of 
48($^{+25}_{-34}$) W.u.  This result represents an unambiguous 
determination of the octupole collectivity, is larger than 
any available theoretical prediction, 
and is consistent with octupole deformation.
\end{abstract}


\maketitle

The concept of spontaneous symmetry breaking in the nuclear density
distribution can be applied to the description of the collective
properties of nuclei \cite{Frauendorf2001}. The coupling between
pairs of nucleons occupying close-lying orbitals with
$\Delta{j}=\Delta{l}=3$ can result in strong octupole correlations,
which can break not only rotational but also reflection symmetry in
the nuclear intrinsic frame \cite{Butler1996}. Nuclei in at least
two regions of the nuclear chart have been identified where both
valence protons and neutrons occupy such orbitals near the Fermi
surface, and they are expected to exhibit signatures of strong
octupole correlations. In fact, the strength of these correlations
can be such that rotational bands with alternating parity appear, 
and these have been commonly interpreted in terms of the
rotation of octupole-deformed nuclei. In the Ra-Th region, recent
measurements of $E3$ transition strengths in $^{220}$Rn and
$^{224}$Ra \cite{Gaffney2013} have validated this interpretation;
and the observed collective structure in $^{224}$Ra is associated
with an octupole shape.

Evidence for octupole collectivity has been inferred in the region
centered around neutron-rich Ba nuclei from $\gamma$-ray studies of
fission fragments \cite{Phillips1986,Urban1997}. Signatures such as
the presence of both $I^+$$\rightarrow$$(I-1)^-$ and
$I^-$$\rightarrow$$(I-1)^+$ enhanced $E1$ transitions linking levels of
the ground-state and negative-parity bands at low and moderate spin have
been reported. These are consistent with expectations of strong
octupole correlations, but whether these are sufficient to stabilize
an octupole shape remains an open question which can be addressed by
measurements of the $E3$ strength. A measurement of the latter
strength is best carried out via sub-barrier Coulomb excitation
\cite{Gaffney2013}, a technique that has only recently become
available for nuclei in the Ba region as it requires the
acceleration of short-lived, radioactive beams.

In this Letter, results from a multi-step Coulomb excitation
experiment with a $^{144}$Ba beam are reported. Besides taking
advantage of new capabilities of acceleration of a radioactive beam,
the measurements also benefited from superior Doppler reconstruction
enabled by the combination of highly-segmented particle counters
with $\gamma$-ray tracking \cite{Paschalis2013} (CHICO2 and GRETINA, 
respectively---see below).

The experiment was conducted at the Argonne Tandem Linac Accelerator System 
(ATLAS).  The $^{144}$Ba beam was produced by the CAlifornium Rare 
Ion Breeder Upgrade (CARIBU) consisting of a $\sim$1.7 Ci 
$^{252}$Cf fission source 
coupled to a He gas catcher capable of thermalizing and extracting the 
fission fragments with high efficiency before filtering them through
an isobar separator \cite{Savard2015,Savard2008}.  
To maximize the extraction of $^{144}$Ba from the system, 
the 2$^+$ charge state was selected for subsequent production of the 
$A$=144 beam.  The latter was charge-bred in 
an Electron Cyclotron Resonance (ECR) ion source to charge state $q$=28$^+$ 
before acceleration through ATLAS.   
Unfortunately, a number of stable contaminants with approximately the same 
$A/q$=5.14, originating from the ECR source, were present with 
the radioactive $A$=144 beam.  These were $^{180}$Hf$^{35+}$, 
$^{134}$Xe$^{26+}$, $^{113}$Cd$^{22+}$, and $^{108}$Cd$^{21+}$ and, 
additionally, $^{36}$Ar$^{7+}$ which was intentionally 
injected into the source as a pilot beam for tuning purposes prior to the 
experiment.  

The 650-MeV $^{144}$Ba beam was passed through a 5~mm diameter 
collimator which was positioned 10.2~cm upstream from a 
1.0~mg/cm$^2$-thick $^{208}$Pb target (99.86\% isotopic purity).  The front 
surface of the target was coated with a 6~$\mu$g/cm$^2$ Al layer and the back 
with 40~$\mu$g/cm$^2$ C. The radioactive beam current 
was monitored with a large HPGe detector positioned just behind 
the beam dump.  The absolute beam intensity was estimated based on the 
yield of the 397-keV $\gamma$ ray emitted following $^{144}$La 
$\beta$ decay ($t_{1/2}$=\,40.8 s \cite{Sonzogni2001}) and determined to be 
8$\times$10$^3$ $^{144}$Ba ions per second.

The experimental setup included the Gamma-Ray Energy Tracking In-beam Nuclear 
Array (GRETINA) \cite{Paschalis2013} for $\gamma$-ray detection 
and CHICO2, a recently upgraded version of the Compact Heavy Ion 
COunter (CHICO) \cite{Simon2000}, for charged-particle detection.  
CHICO2 is characterized by a much-improved $\phi$ (azimuthal) angular 
resolution over that of CHICO.  It is composed of 
20 parallel-plate avalanche counters (PPACs) arranged symmetrically 
around the beam axis.  Each PPAC consists of an aluminized polypropylene 
anode and a pixelated cathode board with a position resolution 
(FWHM) of 1.6$^\circ$ in $\theta$ (polar angle) and 2.5$^\circ$ in 
$\phi$.  The fast anode signal (1.2 ns, FWHM) provides the time difference 
between 2 PPAC events and is used to distinguish between heavy and light 
reaction products, as well as to discriminate between the various beam 
contaminants.  In addition, CHICO2 data provide 
the trajectories of the reaction products required for a precise 
event-by-event Doppler correction of the 
$\gamma$-ray information.   This correction 
also relies on the performance of GRETINA, a spectrometer 
composed of 7 modules, each with 4 segmented HPGe detectors, where the 
segmentation allows for a position resolution of 4.5 mm (FWHM) 
\cite{Paschalis2013} and enables the tracking of multiple interactions 
by a single $\gamma$ ray through the detector.  

\begin{figure*}
    \includegraphics[width=40pc,trim=3mm 0mm 38mm 20mm,clip]
{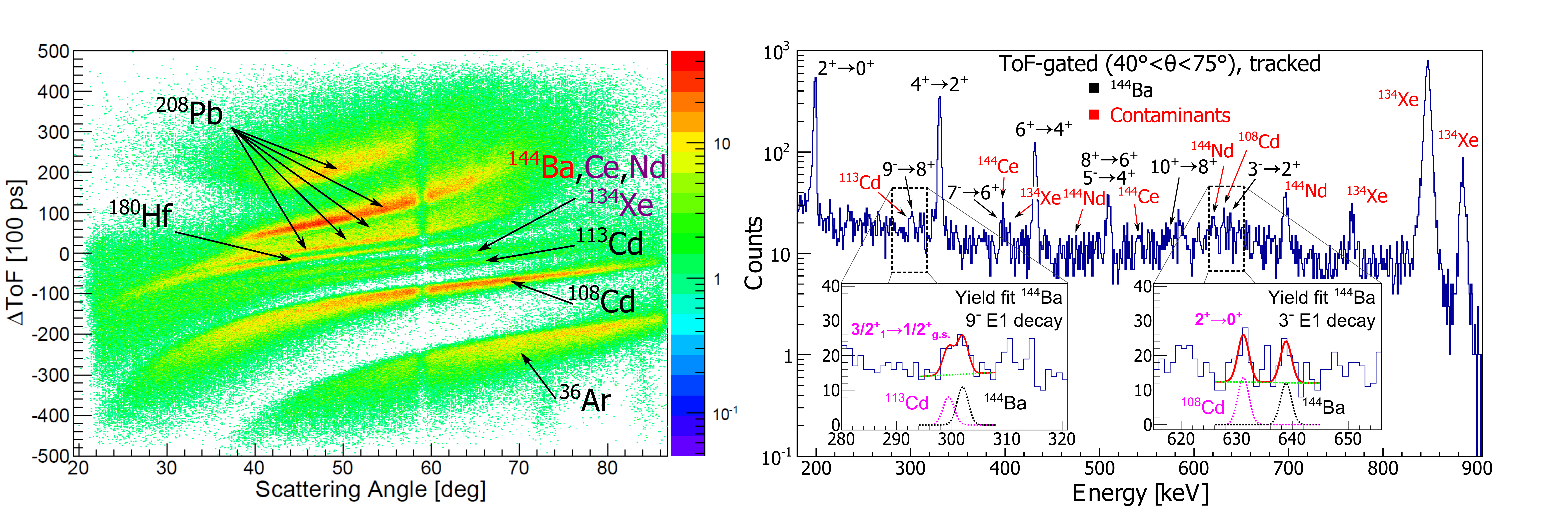}
    \caption{Left: The particle spectrum from CHICO2 measured 
in coincidence with a $\gamma$ ray in GRETINA.  The plot provides the 
difference in ToF between the beam and target nuclei versus the 
scattering angle ($\theta$).  
The various beam contaminants are labeled.  
Right: The $\gamma$-ray spectrum measured in GRETINA gated on the $A$=144 
group in the CHICO2 spectrum (left).  A number of contaminant peaks are 
visible in addition to the $^{144}$Ba $\gamma$ rays.  Note that the 
energy-tracking capabilities of GRETINA have been utilized to help reduce 
the Compton background produced by the high-energy $^{134}$Xe transition.  
Two examples of fits used to extract the yields for $E1$ transitions from 
negative-parity states in $^{144}$Ba are shown in the insets. }
    \label{fig:spectrum}
\end{figure*}

A time-of-flight (ToF) particle spectrum from CHICO2 and the corresponding 
$\gamma$-ray spectrum are 
presented in Fig.~\ref{fig:spectrum}.  The various beam contaminants can be 
identified, and the temporal and spatial 
resolutions are adequate to effectively separate them from the 
$^{144}$Ba beam, except for the $A$=144 isobars and $^{134}$Xe.  
The right side of Fig.~\ref{fig:spectrum} displays the corresponding 
$\gamma$-ray spectrum from GRETINA, with the coincidence requirement of a 
$A$=144 particle detected between 40$^\circ$ and 75$^\circ$.  Clearly, the 
contaminants add significant complexity to the spectrum, particularly 
$^{134}$Xe whose 2$^+$$\rightarrow$0$^+$, 847-keV transition results in a 
significant background contribution under all the $^{144}$Ba $\gamma$~rays 
of interest.  Nevertheless, a number of 
$^{144}$Ba lines have been clearly associated with transitions from 
states with spin as high as 10$\hbar$.  The spectrum includes 
deexcitations from negative-parity levels which are populated in Coulomb 
excitation primarily through $E3$ excitations. 
 
In extracting the yields of the various $\gamma$ rays of interest, 
care was taken to identify all of the nearby contaminants, often through the 
use of additional gates in the ToF spectrum.  
This was especially important for the relatively weak transitions from 
negative-parity states.  The extraction of those yields was further aided by 
prior knowledge of the corresponding $\gamma$-ray energies 
\cite{Sonzogni2001}.  The two insets in 
the spectrum of Fig.~\ref{fig:spectrum} illustrate some of the results for 
the transitions 3$^-$$\rightarrow$2$^+$ (639.0~keV) and 
9$^-$$\rightarrow$8$^+$ (302.1~keV) that were particularly challenging 
because of the presence of contaminants identified in the figure.
The only $\gamma$~ray that could not be individually resolved was the 
5$^-$$\rightarrow$4$^+$ transition at 508.7~keV, close to the more intense 
8$^+$$\rightarrow$6$^+$ one at 509.3~keV 
within $^{144}$Ba itself.  For these two transitions, only the combined 
yield was considered in the Coulomb-excitation analysis.  
Nevertheless, despite the lack of direct decay information from the 
5$^-$ level, the yield data from the 9$^-$ and 7$^-$ states above and the 
3$^-$ level below it provided sufficient information to determine the 
relevant excitation probabilities by the various possible (coupled) channels.

\begin{figure}
    \includegraphics[width=20pc,trim=21mm 43mm 26mm 44mm,clip]
{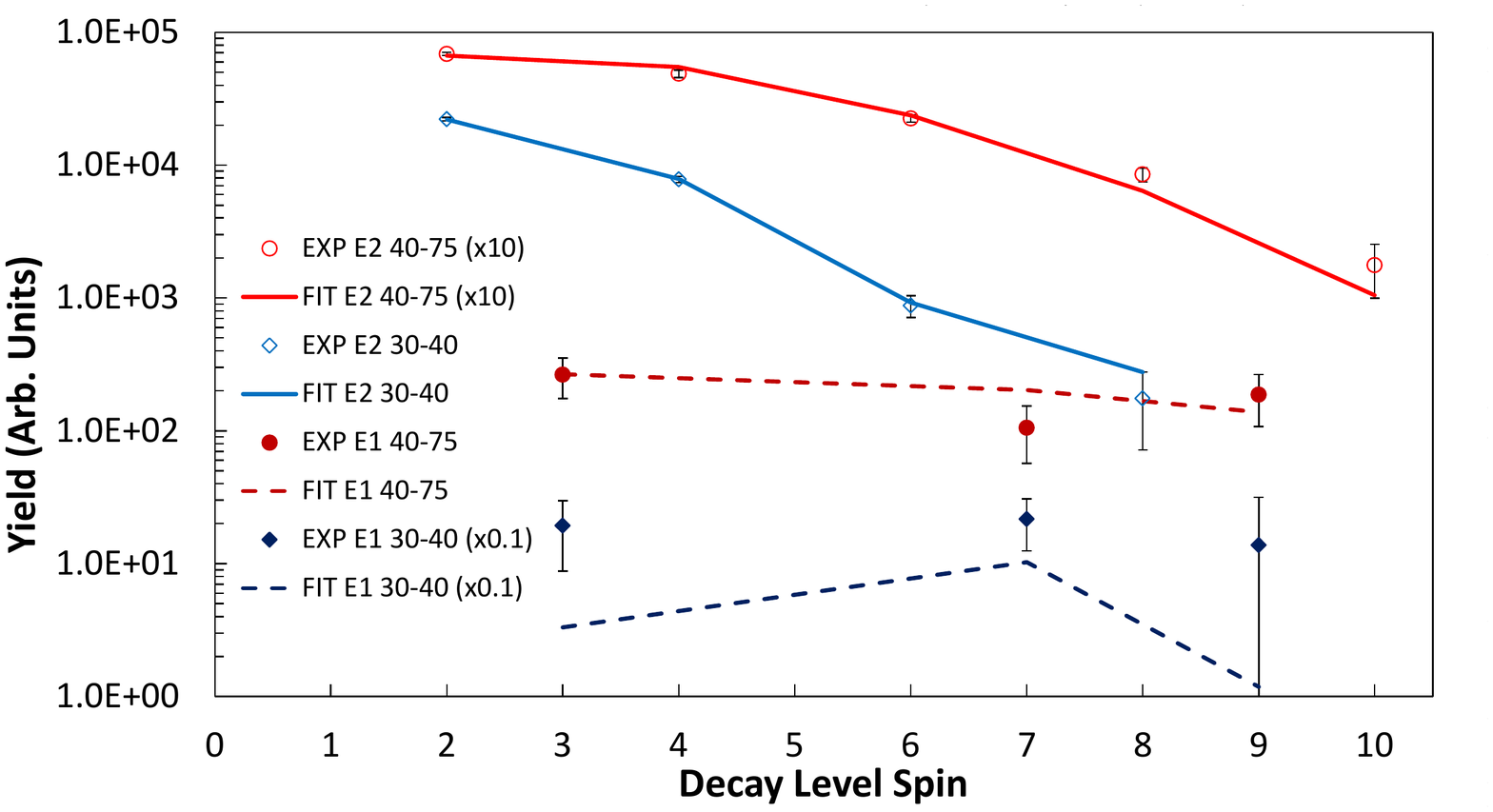}
    \caption{Comparison of the experimental yields 
and uncertainties with those calculated with \textsc{gosia} based on the 
set of matrix elements resulting in 
the best overall agreement with all the available experimental 
data, including previously 
measured lifetimes \cite{Mach1990,Shneidman2005,Biswas2005} and 
branching ratios \cite{Sonzogni2001}.  The 30$^\circ$--40$^\circ$ $E1$ 
and 40$^\circ$--75$^\circ$ $E2$ data 
sets have been renormalized for ease of viewing.  See text for details. }
    \label{fig:yields}
\end{figure}

The $\gamma$-ray detection efficiency was measured with standard $^{182}$Ta, 
$^{152}$Eu, $^{136}$Cs, and $^{60}$Co sources under tracking 
conditions identical to those used in the experiment.  
Intensity ratios between the strongest 
peaks from $^{144}$Ba in the tracked spectrum were verified through 
comparison with those in the corresponding untracked spectrum.
The efficiency-corrected $\gamma$-ray intensities and the associated 
uncertainties can be found in Fig.~\ref{fig:yields}.  For 
the Coulomb-excitation analysis, yields were extracted for two separate 
angular ranges, 30$^\circ$--40$^\circ$ and 40$^\circ$--75$^\circ$ (lab 
frame), as the available statistics did not allow for more restrictive 
intervals.  The intensities for both ranges are displayed in the figure.  
Note that the measured angular distributions are 
such that the $\gamma$-ray yield associated with the $E2$ transition 
deexciting the 10$^+$ state could only be extracted in the 
40$^\circ$--75$^\circ$ gate.  Furthermore, in Fig.~\ref{fig:yields}, the 
data sets measured for $E1$ transitions in the 
30$^\circ$--40$^\circ$ interval 
and for $E2$ $\gamma$ rays in the 40$^\circ$--75$^\circ$ one have been 
renormalized to facilitate their comparative display. 

The experimental yields were analyzed with the semiclassical 
Coulomb-excitation code \textsc{gosia} \cite{Czosnyka1983,*GosiaManualURL} 
which calculates transition intensities for a 
given set of experimental conditions and nuclear matrix elements.  The latter 
are then varied until the set giving the best agreement with the data is 
found, based on a least-squares search.  In the present analysis, states up 
to 14$^+$ in the ground-state band and up to 15$^-$ in the 
negative-parity sequence were considered together 
with the associated $E1$, $E2$, 
and $E3$ matrix elements, totaling 70 in all.  The number of free parameters 
used to fit the limited data set was reduced by coupling the matrix elements 
according to the rigid-rotor prescription 
\cite{Phillips1986,Wollersheim1993}.  Although $^{144}$Ba is considered 
to have moderate deformation only ($\beta_2$$\sim$0.2), 
such a treatment has been validated theoretically for 
even less-deformed cases \cite{Robledo2012}.  Furthermore, in the error 
analysis, the rigid-rotor constraint was released (see discussion 
below).  For the least-squares minimization, 
the constraint requires that the elements 
for a given multipolarity are determined by a single parameter; e.g., the 
intrinsic dipole moment $D_0$ for $E1$ matrix elements, and the quadrupole 
$Q_2$ and octupole $Q_3$ moments for the $E2$ and $E3$ 
elements, respectively.  Here, the $E1$ and $E3$ matrix elements 
were each fit using a single parameter, 
while the elements for the 2$^+$$\rightarrow$4$^+$, 4$^+$$\rightarrow$6$^+$, 
and 6$^+$$\rightarrow$8$^+$ transitions were allowed to vary 
independently, but with constraints provided by lifetime data 
\cite{Mach1990,Shneidman2005,Biswas2005}.  All other $E2$ elements were 
coupled to the 0$^+$$\rightarrow$2$^+$ one, where a precise lifetime is 
available for the 2$^+$ state \cite{Sonzogni2001}.  Available lifetime data 
as well as experimental branching ratios \cite{Sonzogni2001} were also used 
to constrain the $E1$ matrix elements (see below).  
Note that the computations with \textsc{gosia} 
include effects impacting the $\gamma$-ray angular distributions such as 
nuclear de-orientation, relativistic corrections, and detector 
geometry \cite{Czosnyka1983,*GosiaManualURL}.  As an independent check of 
the analysis, \textsc{gosia} was also used to calculate the 
$\gamma$-ray yield 
ratio of the 4$^+$$\rightarrow$2$^+$ and 2$^+$$\rightarrow$0$^+$ transitions 
in $^{134}$Xe, based on the experimental $B(E2)$ probabilities of 
Ref.~\cite{Sonzogni2004}, for scattering angles between 40$^\circ$ and 
75$^\circ$ (same as Fig.~\ref{fig:spectrum}).  The calculated ratio of 0.077 
agrees well with the measured value of 0.078(4), providing added confidence 
in the analysis.

The measured $^{144}$Ba transition yields are compared with the best 
fit results in Fig.~\ref{fig:yields}, while the associated 
$E2$ and $E3$ matrix elements can be 
found in Table~\ref{tab:me}.  Quoted 
errors on the various fit values reflect both the uncertainties associated 
with the data and those originating from correlations between the various fit 
parameters---see Refs.~\cite{Czosnyka1983,*GosiaManualURL,Ibbotson1997} for 
details.  Note that the resulting $E1$ matrix elements are constrained 
primarily by the available data \cite{Sonzogni2001} on branching ratios in 
the decays from the states of interest and display little 
sensitivity to the Coulomb excitation yields.  The $E2$ matrix elements are 
also constrained well by both the available lifetime and branching ratio 
data, and this is reflected in the reported errors in Table~\ref{tab:me}.  
In this context, the matrix element governing the 
2$^+$$\rightarrow$4$^+$ transition deserves some discussion.  Three lifetime 
measurements for the 4$^+$ state are available from the literature.  
Reference~\cite{Mach1990} reports a 49(7) ps mean life 
measured with a fast-timing 
method following $\beta$ decay.  Values of 74(4) ps \cite{Shneidman2005} 
and 71(6) \cite{Biswas2005}, respectively, were determined in 
recoil-distance Doppler-shift 
measurements following $^{252}$Cf fission.  The present analysis results in 
a better overall fit if the shorter lifetime from the decay study is 
used to constrain the fit rather than the larger values obtained 
in the fission studies.  It is 
possible that the two measurements following fission suffer from difficulties 
in properly accounting for feeding into the 4$^+$ level that are absent when 
the state is fed in $\beta$ decay.  All in all, however, the fit results 
are consistent with the available lifetime data for the observed states.

\begin{table}
 \caption{\label{tab:me} The final $E2$ and $E3$ matrix elements 
($e \cdot b^{\lambda /2}$) based on the \textsc{gosia} fit to 
experimental data.}
 \begin{ruledtabular}
  \begin{tabular}{ccc}
$I_i^\pi \rightarrow I_f^\pi$ & $E\lambda$ & 
$\langle I_f^\pi \| \mathcal{M}(E\lambda) \| I_i^\pi \rangle$ \\
\hline
$0^+\rightarrow 2^+$ & $E2$ & 1.042($^{+17}_{-22}$) \\
$2^+\rightarrow 4^+$ & $E2$ & 1.860($^{+86}_{-81}$) \\ 
$4^+\rightarrow 6^+$ & $E2$ & 1.78($^{+12}_{-10}$) \\
$6^+\rightarrow 8^+$ & $E2$ & 2.04($^{+35}_{-23}$) \\
$0^+\rightarrow 3^-$ & $E3$ & 0.65($^{+17}_{-23}$) \\
$2^+\rightarrow 5^-$ & $E3$ & $<1.2$ \\
$4^+\rightarrow 7^-$ & $E3$ & $<1.6$ \\
  \end{tabular}
 \end{ruledtabular}
\end{table}

The main goal of the present measurement was the determination of the $E3$ 
excitation strength in $^{144}$Ba.  Values derived from the fit for the $E3$ 
matrix elements are given in Table~\ref{tab:me}.  While most of these 
elements were not constrained well by the available yields, it was 
nevertheless possible to extract upper limits for the excitations to the 
5$^-$ and 7$^-$ states as well as a value, albeit with sizeable error bars, 
for the 0$^+$$\rightarrow$3$^-$ $E3$ excitation.  The latter value, 
0.65($^{+17}_{-23}$) $e$b$^{3/2}$, corresponds to a reduced transition 
probability $B(E3$;3$^-$$\rightarrow$0$^+$)=48($^{+25}_{-34}$) W.u.  
Note that 
this reported value was obtained under the assumption that the relative 
sign between the sets of electric dipole ($E1$) and octupole ($E3$) matrix 
elements is the same.  In the event that these two sets are of opposite sign, 
the interference term in the calculated excitation probabilities 
\cite{Czosnyka1983,*GosiaManualURL,Wollersheim1993} would 
translate into a reduction in magnitude of $\sim$10\% for the $E3$ matrix 
element of interest; i.e., well within the quoted errors.

The newly measured $E3$ strength can be compared with several theoretical 
values from the literature.  In particular, the $B(E3$;3$^-$
$\rightarrow$0$^+$) probability has been 
calculated using various beyond mean-field approaches 
\cite{Egido1990,Egido1991,Robledo2010}.  The largest predicted value is 
$B(E3)$=20 W.u. \cite{Egido1990}.  
Additionally, and most recently, this $B(E3)$ 
quantity has been calculated with an algebraic approach where a mean-field 
potential energy surface was mapped onto an interacting boson model 
(IBM) Hamiltonian \cite{Nomura2014}.  The latter yielded a similar 
value of $B(E3)$=24 W.u.; the largest strength predicted in 
$^{144}$Ba to date.

Besides the $B(E3)$ probability, the transition octupole moment has 
been calculated using a 
cluster model \cite{Shneidman2005}.  As mentioned earlier, the present 
analysis assumes the rotational limit which implies an intrinsic octupole 
moment with a simple relationship to the transition matrix elements.  The 
present measurement gives a value 
$Q_3$=1.73($^{+45}_{-62}$)$\times$10$^3$ $e$ fm$^3$ compared to 
the prediction in Ref.~\cite{Shneidman2005} of 
1.409$\times$10$^3$ $e$ fm$^3$.  The latter provides the closest 
agreement of any calculated value, although it is based on completely 
different model assumptions.  Nevertheless the error bar on the 
measured value, as mentioned earlier, does include correlations with 
various other matrix elements free from the constraints provided by the 
rigid-rotor assumption, making it essentially model-independent.  
The removal of this constraint in the error analysis is, at least partially, 
responsible for the fact that only upper limits on the 
$\langle 5^- \| \mathcal{M}(E3) \| 2^+ \rangle$ and 
$\langle 7^- \| \mathcal{M}(E3) \| 4^+ \rangle$ matrix elements could be 
determined (Table~\ref{tab:me}).

Going a step further, the octupole moment can be related (with the 
standard assumption of axial symmetry) to the commonly-used 
$\beta_\lambda$ shape parameters 
\cite{Leander1988} describing the nuclear surface as an expansion of the 
spherical harmonics.  Using the quadrupole and octupole moments from the fit, 
a value of 0.17($^{+4}_{-6}$) is derived for the octupole shape parameter 
$\beta_3$ (with $\beta_2$=0.18; the quadrupole moment being largely 
constrained by the measured 2$^+$ lifetime \cite{Sonzogni2001}), under the 
assumption that $\beta_4$ and higher terms in 
the deformation can be neglected.  Generally speaking, such terms 
are expected to deviate significantly from 0 and may play an important 
role in the overall nuclear shape and binding energy 
\cite{Sobiczewski1988,Cwiok1989}, however their relationship 
to the octupole moment is 2nd-order when compared to $\beta_3$ 
\cite{Leander1988}.  Indeed, variations of $\beta_4$ within 
a reasonable range (0 to 0.20), result in a small effect on $\beta_3$ 
($<$10\% for fixed $Q_3$).

The conversion to $\beta_3$ enables comparisons with several additional 
theoretical studies within mean-field approaches \cite
{Nomura2014,Nazarewicz1984,Sobiczewski1988,Moller1995,Zhang2010,Wang2015}.
The largest value is calculated in Ref.~\cite{Moller1995} with 
$|\beta_3|$=0.126.  As 
a matter of fact, Ref.~\cite{Moller1995} presents a comprehensive 
calculation of ground state shapes for 8979 nuclei, covering most of the 
nuclear landscape, up to $A$=339.  The 
measured $\beta_3$ value is larger than any calculated one for nuclei 
with $A$$<$316, although a number of measured $B(E3)$ strengths for 
nuclei with $N$$<$60 have indicated larger $\beta_3$ values 
\cite{Wollersheim1993} than those computed.
Considering the various theoretical calculations of octupole-related 
parameters for $^{144}$Ba, the computed values systematically 
under-predict the present experimental results; the average calculated 
$\beta_3$ deformation between Refs.~\cite
{Nomura2014,Nazarewicz1984,Sobiczewski1988,Moller1995,Zhang2010,Wang2015} 
is less than 0.11, differing from the measured value by more than 
1 standard deviation.  Therefore, generally speaking, 
octupole correlations in $^{144}$Ba are likely stronger than the 
models imply, however the large uncertainty on the present result does not 
allow one to elaborate further.

In conclusion, a number of new developments; i.e., 
the first post-acceleration of a
$^{144}$Ba beam combined with the enhanced performance provided by
particle detection with high angular sensitivity by CHICO2 and the
$\gamma$-ray tracking ability of the GRETINA array proved vital to the 
success of this measurement.  With the determination of the $^{144}$Ba $E3$ 
matrix element 
$\langle 3_1^- \| M(E3) \|0_1^+\rangle=0.65(^{+17}_{-23})$ $e$b$^{3/2}$, 
this measurement provides the first direct 
experimental evidence for significantly enhanced strength of octupole 
correlations in the region centered around neutron-rich Ba nuclei.
Moreover, despite significant uncertainties on the measurement, 
the data also indicate an octupole strength 
larger than calculated in various theoretical approaches.

This work was funded by the U.S. Department of Energy, Office of Science, 
Office of Nuclear Physics, under Contract no. DE-AC02-06CH11357 (ANL), 
DE-AC02-05CH11231 (LBNL, GRETINA), DE-AC52-07NA27344 (LLNL), DOE Grant No. 
DE-FG02-94ER40834 (UM), and the 
National Science Foundation Grant No. PHY-01401574 (FSU) and 
PHY-1068192 (ND).  M.S. and E.T.G. 
were supported by STFC (UK) Grant No. ST/L005808/1.
This research used resources of ANL's ATLAS facility, which is a DOE Office 
of Science User Facility.

\bibliography{BaRefs}

\end{document}